\documentclass[aps,prl,amsmath,amssymb,footinbib,twocolumn,showpacs]{revtex4}
\usepackage{graphicx}
\usepackage{amsmath}
\usepackage{enumitem}
\usepackage{natbib}
\usepackage{float}
\usepackage{verbatim}
\usepackage{bm}
\usepackage[babel=true]{csquotes}
\usepackage[usenames]{color}

\begin{document}

\title{Non-equilibrium spin transport in Zeeman-split superconductors}
\author{Tatiana Krishtop}
\author{Manuel Houzet}
\author{Julia S. Meyer}
\affiliation{Univ.~Grenoble Alpes, INAC-SPSMS, F-38000 Grenoble, France}
\affiliation{CEA, INAC-SPSMS, F-38000 Grenoble, France}
\begin{abstract}
We investigate theoretically the non-local conductance through a superconducting wire in tunnel contact with normal and ferromagnetic leads. In the presence of an in-plane magnetic field, the superconducting density of states is spin-split, and the current injected from the normal lead is spin-polarized. A non-local conductance that is antisymmetric with the applied voltage can be measured with a ferromagnetic lead. It persists for a distance between the contacts that is larger than both the charge-imbalance relaxation length and the normal-state spin relaxation length. We determine its amplitude by considering two extreme models of weak and strong internal equilibration of the superconducting quasiparticles due to electron-electron interactions. We find that the non-local signal, which was measured in recent experiments and discussed as a spin-imbalance effect, can be interpreted alternatively as the signature of a thermoelectric effect.
\end{abstract}



\maketitle

{\it Introduction.-}
Recent progress in the realization of complex ferromagnet/superconductor (F/S) heterostructures has lead to the emergence of \enquote{superconducting spintronics} as a promising field of research.
A fundamental question concerns the ability of superconductors to sustain the flow of a spin-polarized current. A standard way to study spin injection and relaxation in a normal metal consists in contacting it with two ferromagnetic leads, an injector (I) and a detector (D), and measuring the dependence of the non-local conductance, $g_{\rm nl}=\partial I_{\rm D}/\partial V_{\rm I}$, on their separation.
Such a method was also used to measure the spin lifetime in superconductors~\cite{Johnson94,Poli08,Wakamura2014}. 
Moderate variations of the spin lifetime in the superconducting state, with respect to its normal-state value, could be attributed to magnetic impurities~\cite{Poli08} or weak spin-orbit coupling~\cite{Wakamura2014,note}. 
 
It is also possible to inject a spin-polarized current from a non-ferromagnetic, normal (N) metal into a superconducting film subject to an in-plane magnetic field. Indeed, Bogoliubov quasiparticles in superconductors carry a spin $\frac 12$. The spin degeneracy of the superconducting gap for quasiparticles with opposite spins may thus be lifted by the Zeeman effect. At zero temperature, one expects the injected current to be fully spin-polarized if the applied voltage is in the range between the gaps for electrons with opposite spins~\cite{Meservey94}. Based on this effect, recent experiments have established that the non-local conductance measured with a F detector persists over a distance from a N or F injector that is much longer than the spin relaxation length mentioned above~\cite{Hubler12,Quay13,Wolf13,Wolf14}. This method was also applied at finite frequency~\cite{Quay14}.

Such a non-local signal distinguishes itself from the charge-imbalance effect, which already exists in the absence of a magnetic field and for both N injector and detector \cite{Tinkham72,Tinkham1972a}, by a longer decay length. Moreover, in the case of a N injector, the field induced non-local conductance is antisymmetric with respect to the applied voltage, while the charge-imbalance signal is symmetric. By analogy, it was discussed as a \enquote{spin-imbalance} effect~\cite{Zhao95}. Various scenarios for the relaxation of the signal were discussed, without a clear conclusion.

The aim of our work is to show theoretically that such a signal appears naturally in models where the spin-imbalance -- defined as a difference between distribution functions for quasiparticles with opposite spins -- is absent. Indeed, one needs to distinguish between spin polarization, arising from a density of states effect, and spin imbalance, characterizing unequal occupations. The current injected in the superconductor is accompanied by an energy flow that produces an out-of equilibrium distribution of quasiparticles in the S wire. 
This in turn results in an induced current at the detector, provided that the quasiparticles with opposite spins have different densities of states, such that the superconductor acquires a finite spin polarization, and different probabilities to tunnel through the tunnel barrier. Those conditions are naturally realized, if the superconducting density of states is Zeeman-split and the detector is ferromagnetic.  The detector current then corresponds to the thermoelectric effect predicted in a single S/F tunnel junction in the presence of a magnetic field~\cite{Machon2013,Ozaeta14}.   
        Within this approach, the scale over which the non-local signal disappears is related with energy relaxation, which is set by electron-phonon interactions. Namely, as long as energy cannot relax, the superconductor will carry excess quasiparticles that lead to a non-local signal. In particular, while electron-electron interactions lead to internal equilibartion of the quasiparticles, they do not relax energy and, therefore, do not suppress the signal. We find, however, that they qualitatively change its voltage dependence.

\begin{figure}
\includegraphics[width=0.7\columnwidth]{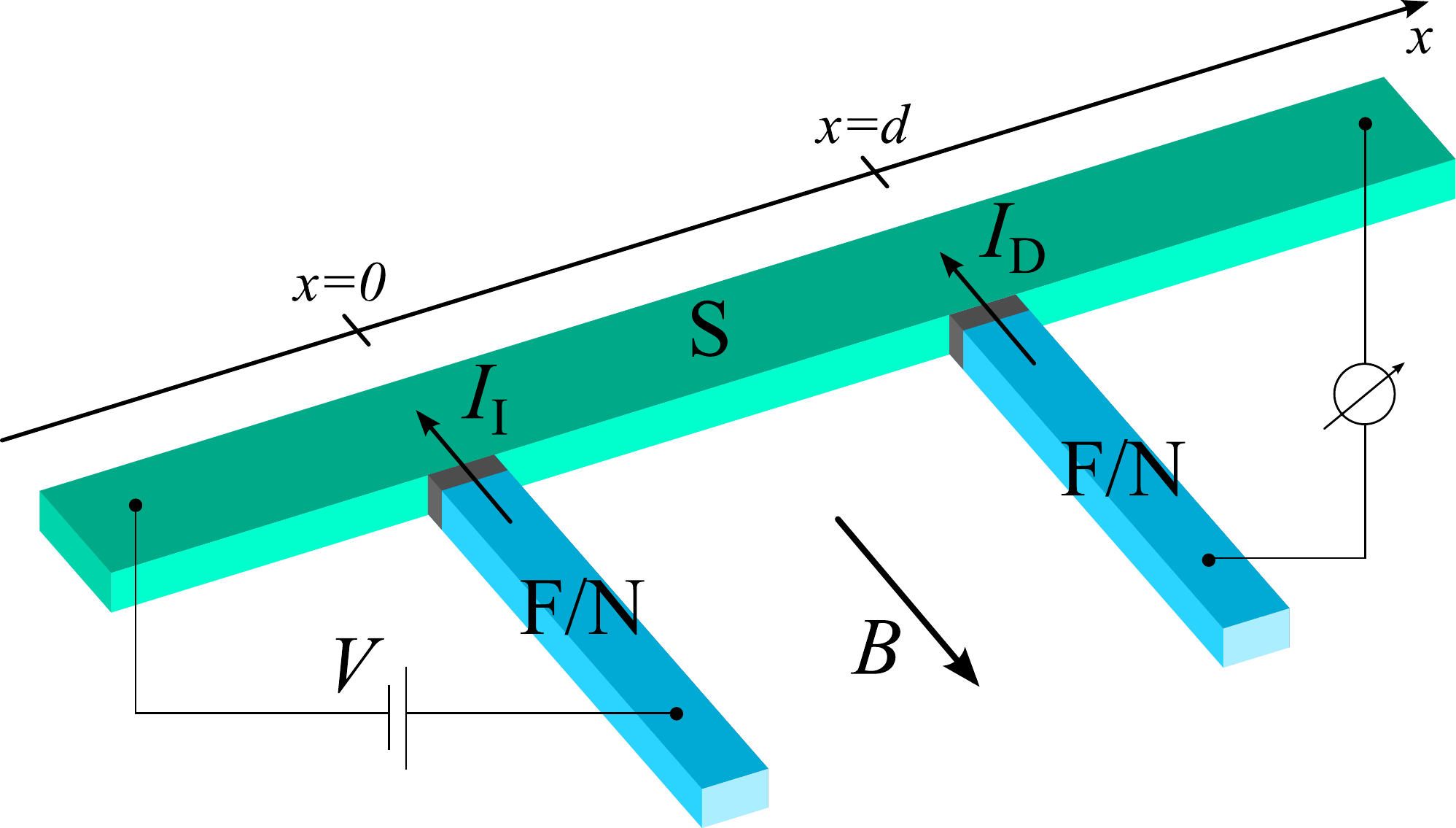}
\caption{\label{F:setup} Setup of the junction.}
\end{figure}

{\it Formalism.-}
The setup that we address consists of a superconducting wire in tunnel contact with two normal or ferromagnetic leads separated by a distance $d$, see Fig.~\ref{F:setup}. The first lead is voltage biased with respect to the superconductor; it acts as a quasiparticle injector. As shown in experiments~\cite{Hubler12,Quay13,Wolf13}, it can result in a detector current flowing through the second contact.
Below we obtain the dependence of the injector and detector currents, $I_{\rm I}$ and $I_{\rm D}$, respectively, on the voltage bias $V$, both in the regimes of slow and fast electron-electron relaxation. While the formalism is more general, we specify the results only for a N/S/F structure, i.e.,  a N injector and F detector.

Using the tunnel Hamiltonian theory of Ref.~\cite{Tinkham1972a}, we find that the current flowing between a normal or ferromagnetic lead at equilibrium and a superconductor with an out-of-equilibrium quasiparticle distribution is given by 
\begin{widetext}
\begin{eqnarray}
\label{courant-tunnel}
I_{\alpha}&=&\frac{G_{\alpha}}{e}\sum_\sigma\int_\Delta^\infty dE\,\Big\{
\nu(E) \left[\frac{f_0(E_{\sigma}+eV_\alpha)-f_0(E_{\sigma}-eV_\alpha)}2+ \sigma P_\alpha\left(f_0
(E_\sigma)-
\frac{f_0(E_{\sigma}+eV_\alpha)+f_0(E_{\sigma}-eV_\alpha)}2\right)\right]\nonumber\\
&&\qquad\qquad\qquad\qquad+ \sigma P_\alpha \nu(E)\left[f^+_\sigma(E_\sigma)-f_0(E_\sigma)\right]
+ f^-_\sigma(E_\sigma)
\Big\}.
\end{eqnarray}
\end{widetext}
Here, $f_0$ is the Fermi distribution at temperature $T_0$ in the leads, $V_\alpha$ is the voltage bias between lead $\alpha={\rm I}, {\rm D}$ and the superconductor, with $V_{\rm I}=V$ and $V_{\rm D}=0$, 
$G_{\alpha}$ is the normal-state tunnel conductance, $P_\alpha$ is the spin-polarization of the lead ($|P_\alpha|<1$ and $P_\alpha=0$ for a normal lead), $\Delta$ is the superconducting gap, $\nu(E)=E/\sqrt{E^2-\Delta^2}$ is the reduced BCS density of states, and $E_\sigma=E+\sigma h$, where $h=\mu_B B$ is a Zeeman field and $\sigma=\pm$ (for $\uparrow/\downarrow$) is a spin label. 

The non-local signal is described by the second line of Eq.~\eqref{courant-tunnel}. The out-of-equilibrium distribution functions in the superconductor, $f^\pm_\sigma(E)=[f^>_\sigma(E)\pm f^<_\sigma(E)]/2$, 
are related with the distributions for electron-like $[f^>_\sigma(E)]$ and hole-like $[f^<_\sigma(E)]$ Bogoliubov excitations with spin $\sigma$ and energy $E>\Delta+\sigma h$, respectively.
The last term in Eq.~\eqref{courant-tunnel} corresponds to a charge-imbalance contribution to the current~\cite{Tinkham1972a}. Deviations of $f^+_\sigma$ from its equilibrium value, $f_0$, yield a contribution to the current proportional to $P_\alpha$, which therefore exists only when the lead is ferromagnetic.

To evaluate the current \eqref{courant-tunnel}, we need to determine the superconducting gap and distribution functions in the vicinity of the tunnel junctions. The former should satisfy the local self-consistency equation
\begin{equation} 
\label{eq:self}
1=\lambda \int_\Delta^\Omega \frac{dE}{\sqrt{E^2-\Delta^2}} \left[1- \sum_\sigma f^+_\sigma(E_\sigma)\right].
\end{equation}
Here, the pairing constant $\lambda$ and Debye frequency $\Omega$ are related with the BCS gap at zero temperature through $\Delta_0=2\Omega e^{-1/\lambda}$.
To obtain the distribution functions, we may generalize the Boltzmann equation approach for Bogoliubov quasiparticles~\cite{Bardeen59,Aronov1986} 
to Zeeman-split superconductors. In the diffusive regime ($\Delta\tau\ll 1$, where $\tau$ is the elastic scattering time)~\cite{note-Boltzmann}, we find that the distribution functions solve the kinetic equations
\begin{eqnarray} 
\label{eq:kinetic}
-D_\sigma(E)\nabla^2 f^\gtrless_\sigma(E)&=&
{\cal I}^\gtrless_\sigma(E).
\end{eqnarray}
Here, $D_\sigma(E)=D_N/\nu(E_{\bar \sigma})$, where $D_N$  is the diffusion constant in normal state and $\bar\sigma=-\sigma$. The collision integral ${\cal I}^\gtrless_\sigma(E)$ contains elastic and inelastic contributions as well as contributions due to the tunnel injection of quasiparticles from the leads.

The elastic contribution stems from the effect of the in-plane magnetic field on the orbital motion of the electrons as well as from the electron scattering by magnetic impurity and by normal impurities having a large spin-orbit scattering potential.
It takes the form
\begin{eqnarray}
\label{coll-el}
{\cal I}^\gtrless_{\rm{el},\sigma}(E)&=&
\frac 1{\tau_{1\sigma}(E)}\left[f^\lessgtr_\sigma(E)-f^\gtrless_{\sigma}(E)\right]
\nonumber\\
&&
+\frac 1{\tau_{2\sigma}(E)}\left[f^\gtrless_{\bar\sigma}(E)-f^\gtrless_{\sigma}(E)\right]
\nonumber\\
&&
+\frac 1{\tau_{3\sigma}(E)}\left[f^\lessgtr_{\bar\sigma}(E)-f^\gtrless_{\sigma}(E)\right]
,
\end{eqnarray}
where
\begin{subequations}
\begin{eqnarray}
\frac 1{\tau_{1\sigma}(E)}
&=&\left(\frac 1{\tau_{\rm{orb}}}+\frac 1{\tau_{\rm{m}}}\right)\frac{\Delta^2}{E_{\bar\sigma}\xi_{\bar\sigma}},
\\
\frac 1{\tau_{2\sigma}(E)}&=&\frac{\xi_{\bar\sigma}}{E_{\bar\sigma}}
\left(
\frac 1{\tau_{\rm{m}}}\mu_{++}(E)
+\frac 1{\tau_{\rm{so}}}\mu_{+-}(E)
\right),
\\
\frac 1{\tau_{3\sigma}(E)}&=&
\frac{\xi_{\bar\sigma}}{E_{\bar\sigma}}\left(\frac 1{\tau_{\rm{m}}}\mu_{-+}(E)
+
\frac 1{\tau_{\rm{so}}}\mu_{--}(E)\right),
\end{eqnarray}
\end{subequations}
and 
$\mu_{ss'}(E)=\theta(E_\sigma-\Delta)[E_{\bar\sigma}E_\sigma+s\xi_{\bar\sigma}\xi_\sigma+s'\Delta^2]/(\xi_{\bar\sigma}\xi_\sigma)$
for $s,s'=\pm$.
Here, $1/\tau_{\rm{orb}}=D_N(ewB)^2/6$ for an in-plane magnetic field and a wire with a thickness $w\ll\xi_S$, where $\xi_S=\sqrt{D_N/\Delta_0}$ is the superconducting coherence length, whereas $1/\tau_{\rm{m}}$ and $1/\tau_{\rm{so}}$ are the spin-flip rates due to magnetic and spin-orbit impurities, respectively, in the normal state.  These rates will be assumed smaller than $\Delta$, so that they affect the kinetic equations, but not the superconducting density of states. The rate $1/\tau_{1\sigma}$ leads to charge-imbalance relaxation while the rate $1/\tau_{2\sigma}$ leads to spin-imbalance relaxation, whereas the rate $1/\tau_{3\sigma}$ yields both.
Note that the two last terms in the r.h.s. of Eq.~\eqref{coll-el} are only effective when the bands for Bogoliubov quasiparticles with opposite spins overlap, enabling elastic spin relaxation.

The tunnel injection of quasiparticles at position $x_\alpha$ results in the term
\begin{eqnarray}
\label{Itun}
{\cal I}^\gtrless_{\alpha,\sigma}(E)&=&\frac{\delta(x-x_\alpha)}{2e^2\nu_N\Sigma}
G_{\alpha}\sum_{s=\pm}
\frac{1+s\sigma P_\alpha}2
\times\\
&&\quad\times\left(1\pm s\frac{\xi_{\bar\sigma}}{E_{\bar\sigma}}\right)
\left[f_0(E+seV_\alpha)-f^\gtrless_{\sigma}(E)\right].\nonumber
\end{eqnarray}
Here $\nu_N$ is the density of the superconductor (per spin) in the normal state, and  $\Sigma$ is the cross section of the superconducting wire

Finally, the inelastic collision integral contains both electron-electron and electron-phonon contributions, characterized by rates $\gamma_{\rm e-e}$ and $\gamma_{\rm e-ph}$, respectively.  It is important to note that electron-electron interactions do not lead to energy relaxation as discussed in more detail later.

Recent experiments with N/S/N structures demonstrated that, at low temperatures, the relaxation of the charge-imbalance signal under magnetic field was dominated by the orbital depairing mechanism~\cite{Hubler2010,Kleine2010}. The non-local spin signal of Refs.~\cite{Hubler12,Wolf13} persists over a much longer length than the one determined by this effect. Taking this into account, 
we decompose Eq.~\eqref{eq:kinetic} into two diffusion equations for $f^+_\sigma(E)$ and $f^-_\sigma(E)$. 
Assuming that charge imbalance relaxes fast, $1/\tau_{\rm{orb}}\to\infty$, we find $f^-_\sigma(E)=0$. As a consequence, Eq.~\eqref{eq:kinetic} reduces to a single kinetic equation,
\begin{equation}
\label{eq:kinetic2}
-D_\sigma(E)\nabla^2 f^+_\sigma(E)={\cal I}_{\sigma}(E).
\end{equation}
In particular, the elastic contribution to the collision integral reduces to a spin-flip term
\begin{eqnarray}
\label{coll-el2}
{\cal I}_{\rm{el},\sigma}(E)&=&
\frac 1{\tau_{{\rm s}\sigma}(E)}\left[f^+_{\bar\sigma}(E)-f^+_{\sigma}(E)\right]
,
\end{eqnarray}
with elastic spin relaxation rate ${\tau^{-1}_{{\rm s}\sigma}(E)}={\tau^{-1}_{2\sigma}(E)}+{\tau^{-1}_{3\sigma}(E)}$, in agreement with Refs.~\cite{Zhao95,Yamashita2002,Morten05} at $h=0$.

Taking into account that $P_{\rm I}=0$ and $V_{\rm D} =0$ for the N/S/F setup that we consider, Eq.~\eqref{courant-tunnel} reduces to
\begin{equation}
\label{Iinj}
I_{\rm I}=\frac{G_{\rm I}}{2e}\!\sum_\sigma\!\int_\Delta^\infty \!\!\!dE\;
\nu(E)
\left[
f_0(E_{\sigma}+eV)-f_0(E_{\sigma}-eV)
\right]
\end{equation}
for the injector current and
\begin{equation}
\label{Idet}
I_{\rm D}=\frac{P_{\rm D}G_{\rm D}}{e}\sum_\sigma\sigma \int_\Delta^\infty \!\!\!dE\;
\nu(E)
\left[f^+_\sigma(E_{\sigma})-f_0(E_{\sigma})\right]
\end{equation}
for the detector current, respectively. 
Eq.~\eqref{Idet} then yields the non-local conductance $g_{\rm nl}=\partial I_{\rm D}/\partial V_{\rm I}$.

While electron-phonon scattering in superconductors was investigated in detail~\cite{Kaplan76}, much less is known about electron-electron scattering. It was argued recently that, due to the different energy dependence of the two rates, a large or small ratio $\gamma_{\rm e-e}/ \gamma_{\rm e-ph}$ could be achieved in Aluminum (the superconductor used in all cited experiments), depending on the  energy range probed~\cite{Kopnin2009}. To proceed further, we will therefore evaluate the distribution $f^+_\sigma$ that enters Eq.~\eqref{Idet}  in two limiting cases, $\gamma_{\rm  e-e}\ll \gamma_{\rm  e-ph}$ and  $\gamma_{\rm e-e}\gg \gamma_{\rm e-ph}$, and discuss qualitatively the differences in the result for the non-local conductance. Electron-phonon processes will be described phenomenologically by imposing that $f_\sigma^+(E)=f_0(E)$ at a distance $L_\star$ from the tunnel junctions. Characteristic length scales for Al are in the range of several $\mu$m, as discussed, e.g., in Ref.~\cite{Hubler2010}.

{\it Slow internal equilibration.-}
We first assume that electron-electron collisions are very rare. In that case, spin relaxation is  dominated by elastic spin-flip processes. Eq.~\eqref{eq:kinetic2}  may be solved to obtain the distribution function $f_\sigma^+$, both for slow and fast elastic spin relaxation. 

Neglecting elastic spin-flip processes, $1/{\tau_{{\rm s}\sigma}(E)}\to0$, and using the boundary conditions $f_\sigma^+(E;x=-L_\star)=f_\sigma^+(E;x=d+L_\star)=f_0(E)$, we obtain the solution to Eq.~\eqref{eq:kinetic2} at the position of the detector ($x=d$) as
\begin{widetext}
\begin{equation}
\label{eq:fs}
f^+_\sigma(E)
=
f_0(E)
+
\frac{\nu_{{\rm I}}(E_{\bar \sigma})G_{\rm I}
\left[
\frac 12\left(f_0(E+eV)+f_0(E-eV)\right)-f_0(E)
\right]}
{2G_\star+\nu_{{\rm I}}(E_{\bar \sigma})G_{\rm I}+\nu_{{\rm D}}(E_{\bar \sigma})G_{\rm D}
+
({d}/{L_\star})
\left[ G_\star+\nu_{{\rm I}}(E_{\bar \sigma})G_{\rm I}\right]
\left[ G_\star+\nu_{{\rm D}}(E_{\bar \sigma})G_{\rm D}\right]/ G_\star
}
,
\end{equation}
\end{widetext}
where $G_\star=\sigma_N\Sigma/L_\star$ is an effective conductance associated with the electron-phonon relaxation length and $\sigma_N=2e^2\nu_N D_N$ is the normal-state conductivity of the superconducting wire. Here we added a subscript $\alpha$ to the reduced density of states ($\nu_{\alpha}$) to indicate that is should be evaluated with the self-consistent gap $\Delta_\alpha$ at the position $x_\alpha$ of the corresponding lead. Note that, at $G_\star\ll G_{\rm I},G_{\rm D}$ and $d\ll L_\star$, the distribution function does not depend on spin.

As elastic spin-flip processes require spin-up and spin-down states to be available, they modify the distribution function only in the energy range $E>\Delta+h$, where the density of states in both spin bands is finite. Thus, in the opposite regime of fast elastic spin relaxation, $1/{\tau_{{\rm s}\sigma}(E)}\to\infty$, the result \eqref{eq:fs} still holds at energies $\Delta-h<E<\Delta+h$. At $E>\Delta+h$, spin relaxation equilibrates the distributions between up and down spins, $f_\uparrow^+(E)=f_\downarrow^+(E)$. The resulting distribution is given by an equation similar to \eqref{eq:fs} with $\nu_{\alpha}(E_{\bar \sigma})$ replaced by $\bar\nu_\alpha(E_{\bar\sigma})=\sum_\sigma\nu_\alpha(E_{\bar\sigma})/2$.  

\begin{figure}
(a)\!\!\!\!\!\!\!\!\!
\includegraphics[width=0.49\columnwidth]{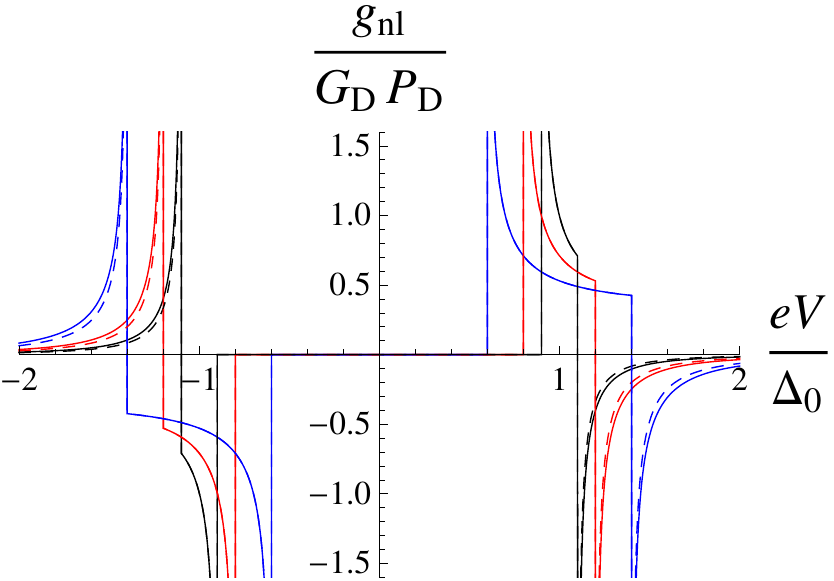}
\hfill
(b)\!\!\!\!\!\!\!\!\!
\includegraphics[width=0.49\columnwidth]{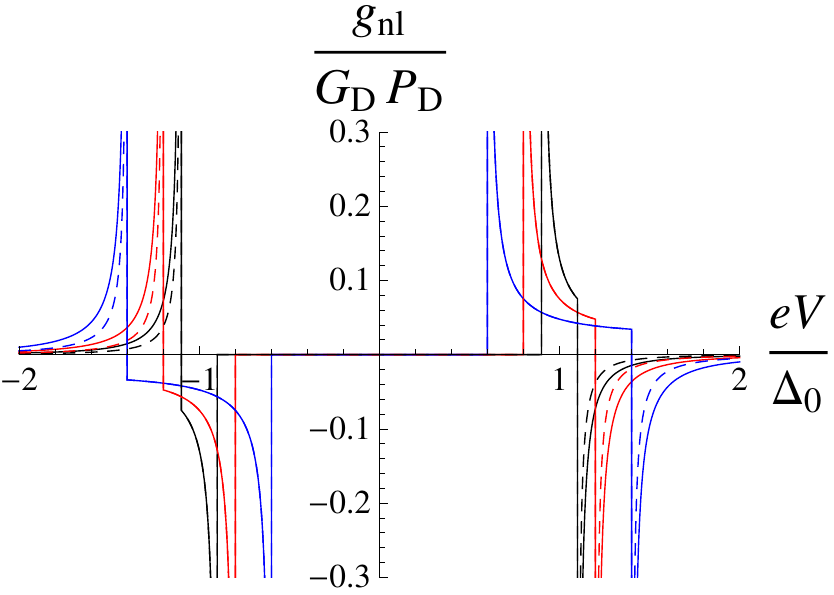}
\caption{\label{F:elastic} Non-local conductance as a function of the bias voltage in the regimes of slow (solid lines) and fast (dashed lines) elastic spin relaxation. Curves are shown for three different Zeeman fields, $h/\Delta_0=0.1$ (black), $0.2$ (red), and $0.4$ (blue). The two figures correspond to $G_\star=0.5G_{\rm D}$ (a) and $G_\star=20G_{\rm D}$ (b). Furthermore, $d\ll L_\star$ and $G_{\rm I}=G_{\rm D}$.}
\end{figure}

The non-local conductance is obtained by 
inserting the distribution \eqref{eq:fs} into Eq.~\eqref{Idet}.  Fig.~\ref{F:elastic} shows results in both regimes of slow and fast elastic spin relaxation. (Note that here we disregarded the self-consistency of the gap.) We observe that both regimes give rise to the same qualitative behavior. Namely, the non-local conductance is the sum of two contributions of opposite signs that have different threshold voltages $(\Delta\pm h)/e$, respectively. At $G_\star\ll G_{\rm I},G_{\rm D}$, spin relaxation has no effect, as $f_\uparrow^+(E)\approx f_\downarrow^+(E)$ in the entire energy range even in the absence of spin-flip processes. At $G_\star\gg G_{\rm I},G_{\rm D}$, elastic spin relaxation leads to a reduction of the peak at the threshold voltage $(\Delta+h)/e$. However, also the overall magnitude of the signal decreases with increasing electron-phonon relaxation, corresponding to increasing $G_\star$. Furthermore, as a function of the separation between injector and detector, the non-local conductance decreases algebraically on the scale of the electron-phonon relaxation length. Therefore, the measured signal is mainly governed by energy relaxation. It is difficult to make a clear distinction between the cases where spin-imbalance is present or absent. 

{\it Fast internal equilibration.-}
In the opposite regime of frequent electron-electron collisions, 
the energy redistribution among Bogoliubov quasiparticles tends to establish a common Fermi distribution $f(E)$, with a local temperature $T$ that may differ from the temperature in the leads, $T_0$. Such heating effects were previously addressed in the context of S/N/S junctions~\cite{Kolenda13}. As the internal equilibration in this so-called \enquote{hot} quasiparticle regime facilitates spin relaxation, we will concentrate on the regime of fast spin relaxation. Thus, there is no spin imbalance. Applying the already mentioned condition of energy conservation
for electron-electron scattering processes, we then obtain the heat equation
\begin{equation}
\label{eq:heat}
-\nabla^2\left[\frac{\sigma_N}{e^2}\sum_\sigma \int_\Delta^\infty \!\!\! dE\; E_\sigma f(E_\sigma)\right] =\sum_{\alpha={\rm I,D}}\dot Q_{\alpha},
\end{equation}
where 
\begin{equation}
\dot Q_{\alpha}=
\frac{G_{\alpha}\delta(x-x_\alpha)}{e^2\Sigma}\sum_\sigma\int_\Delta^\infty \!\!\! dE\;
\nu(E)E_\sigma\left[f_\alpha(E_\sigma)
- f(E_{\sigma})
\right]
\end{equation}
is the power injected at the tunnel junctions and $f_\alpha(E)=[f_0(E+eV_\alpha)+f_0(E-eV_\alpha)]/2$~\cite{conductivity}.

Integrating Eq.~\eqref{eq:heat} along the S wire for nearby injector and detector junctions ($d\ll L_\star$), we find that the local temperature and superconducting gap in the vicinity of the junctions are determined by
\begin{eqnarray}
\label{eq:selfT}
0&=&\sum_\sigma\int_\Delta^\infty \!\!\!dE\,E_\sigma
\Big\{
2G_\star\left[f(E_\sigma)
-\theta(E\!-\!\Delta(T_0))f_0(E_\sigma)\right]
\nonumber\\
&&\qquad\quad\qquad+\nu(E)\sum_\alpha G_\alpha \left[f(E_\sigma)-
f_\alpha(E_\sigma)\right]
\Big\},\quad
\end{eqnarray}
together with the self-consistency equation \eqref{eq:self} that yields $\Delta(T,h)$~\cite{Sarma1963}. The generalization to a finite separation $d$ between injector and detector is straightforward~\cite{note-finite-distance}.

The dependence of the local temperature on the applied voltage shows two thresholds at $eV=\Delta\pm h$, corresponding to the opening of each of the spin channels. For a critical voltage 
\begin{equation}
eV_c=
\sqrt{\frac{\pi^2}3\frac{G_\Sigma+2G_\star}{G_{\rm I}}T_c^2(h)-\frac{2G_\star}{G_{\rm I}} h^2}\,,
\end{equation}
where $G_\Sigma=G_{\rm I}+G_{\rm D}$ and 
$T_c(h)$ is the superconducting critical temperature, the local temperature reaches the critical temperature and, thus, superconductivity is suppressed locally~\cite{transition}. When $G_\star\gg G_{\rm I},G_{\rm D}$, the critical voltage is very large and the local suppression of superconductivity in the voltage range  up to a few $\Delta_0$ is negligible.

\begin{figure}
(a)\!\!\!\!\!\!\!\!\!
\includegraphics[width=0.49\columnwidth]{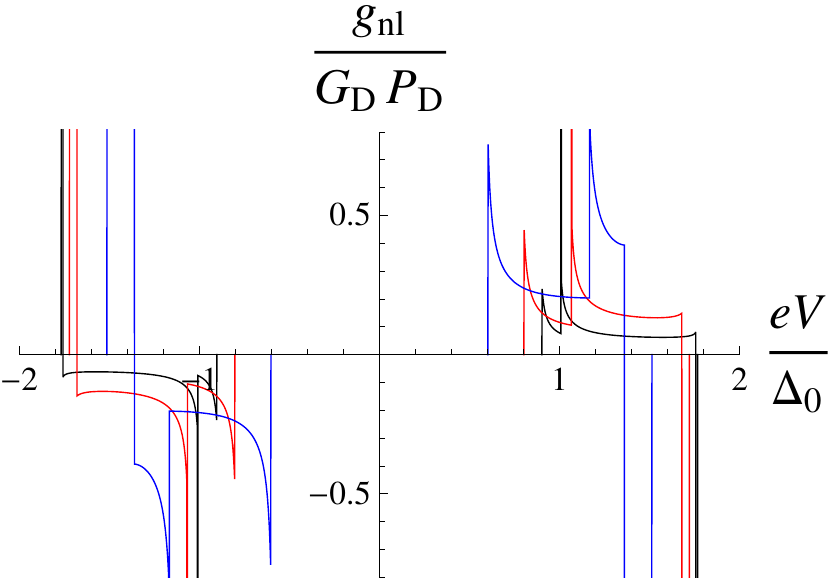}
\hfill
(b)\!\!\!\!\!\!\!\!\!
\includegraphics[width=0.49\columnwidth]{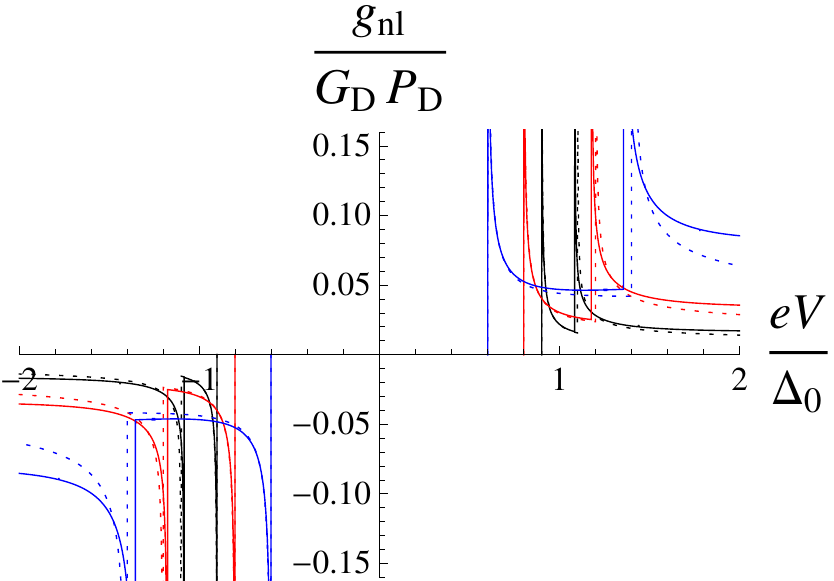}
\caption{\label{fig-inelastic} Non-local conductance as a function of the bias voltage in the \enquote{hot} quasiparticle regime. Same parameters as in Fig.~\ref{F:elastic}. Note that in figure (a), local superconductivity is suppressed completely beyond the critical voltages $eV_c/\Delta_0\approx1.51,1.72,1.77$ for $h/\Delta_0=0.1,0.2,0.4$, respectively. In figure (b), corresponding to a larger value of $G_\star$, the critical voltages are much larger. Thus, in the voltage range shown, the local suppression of superconductivity is small. The results without taking it into account are shown for comparison (dotted lines).}
\end{figure}

The non-local signal exists only for voltages $\Delta_0(h)-h<eV<eV_c$. The current increases with voltage as long as the local density of states remains gapped and then decreases once the local temperature reaches the value for which $\Delta(T,h)=h$ and the gap closes locally. The successive openings of the two spin channels for transport thus yield two peaks of the same sign in the non-local conductance, whereas the gap closing at $V\lesssim V_c$ results in a peak with opposite sign. The position of this last peak beyond which the signal quickly vanishes is determined mainly by the efficiency of energy relaxation, characterized by $G_\star$. Its magnitude may be large because the gapless region is narrow. The results are illustrated in Fig.~\ref{fig-inelastic}.

{\it Discussion.-} We showed that a finite non-local signal exists in the absence of spin imbalance, both for slow and fast electron-electron relaxation. The amplitude of the signal decreases with increasing energy relaxation due to electron-phonon processes. On the other hand, at very weak energy relaxation, the system quickly overheats and superconductivity becomes locally suppressed, leading to a complete suppression of the non-local signal beyond a critical voltage $V_c$. Electron-electron relaxation qualitatively modifies the voltage dependence of the non-local conductance. In particular, if $V_c\gg\Delta_0$, one finds two peaks with {\em opposite} signs at slow electron-electron relaxation whereas one finds two peaks with the {\em same} sign at fast electron-electron relaxation. While spin imbalance may be present, it does not lead to any easily identifiable features in the non-local conductance. A double peak structure in agreement with the scenario of fast electron-electron relaxation is clearly visible in Ref~\cite{Wolf14}. 

{\it Note added.-} In the last stage of this work, we learned of two preprints that address the same effect~\cite{Silaev14,Bobkova14}. 
They use a quasiclassical theory, which is complementary to our kinetic theory and which allows them to incorporate the orbital depairing effect of the magnetic field. While Ref.~\cite{Silaev14} also attributes the signal measured in Refs.~\cite{Hubler12,Quay13} to a thermoelectric effect, none of these preprints consider the \enquote{hot} quasiparticle regime and the local suppression of superconductivity that we discuss here.

\begin{acknowledgments}
We thank M. Aprili, D. Beckmann, and C. Quay for useful discussions. This work is supported by ANR grants ANR-11-JS04-003-01 and ANR-12-BS04-0016-03, and by EU-FP7 Marie Curie IRG.
\end{acknowledgments}


\begin{thebibliography}{99}
\bibitem{note}
Note however that a huge enhancement of the spin lifetime was obtained from a different magnetoresistance measurement on a two-terminal F/S/F junction and was attributed to strong spin-orbit coupling~\cite{Yang10}.
\bibitem{note-finite-distance}
As temperature varies on a scale much larger than the coherence length, it would require to solve two coupled equations similar to Eq.~\eqref{eq:selfT} together with the local self-consistency equations, both at the injector and detector positions.
\end{thebibliography}
\end{document}